\documentclass[12pt,twoside]{article}
\usepackage{correl}
\def\TheTitle{Variational Monte Carlo for the Hubbard model}
\def\TheHeading{VMC for Hubbard}
\def\TheAuthor{Luca F. Tocchio}
\def\TheInstitute{Politecnico di Torino}
\def\TheAddress{Corso Duca degli Abruzzi, 24 10129-Turin, Italy}
\def\TheChapter{}           

\usepackage{mathtools}
\DeclarePairedDelimiter\ket{\lvert}{\rangle}
\DeclarePairedDelimiter\bra{\langle}{\lvert}
\newcommand{\dagga}{\phantom{\dagger}}

\begin{document}
\MakeTitle
\section{Introduction}

One of the reasons why the study of correlated Hamiltonians in two dimensions is one of the most active and debated fields in modern condensed matter physics is the absence of a clear-cut method for tackling non-exactly solvable Hamiltonians, like the Hubbard model. Variational Monte Carlo (VMC) is one of the few methods that allows us to address Hubbard models, including its multi-orbital extensions, directly in two dimensions, describing non-local correlations and ordering, including non-local superconductivity~\cite{Becca2017}. We remark that VMC is a nonperturbative approach which performances are not significantly affected by the dimensionality of the system or by the size of the local Hilbert space. In this respect, VMC might be more efficient than other methods, like density-matrix renormalization group or tensor-network methods, in which the complexity increases with both the range of the interaction and the dimension of the local Hilbert space~\cite{Schollwoeck2011}. It can also provide complementary information with respect to methods like Dynamical Mean-Field Theory~\cite{georges1996}, which can capture accurate dynamical local correlations, but do not include nonlocal spatial correlations beyond mean field.

The VMC methods rely on two main ingredients. The first one is the construction of a variational wave function that approximates the correct ground-state one. This approach relies on the variational principle, stating that the energy of a given quantum state is always bounded from below by the exact ground-state energy, thus giving us guidance to obtain the best solution to the problem. Moreover, a suitable parametrization of the ground state allows us to consider a wide range of different quantum phases (superconductivity, magnetic and charge order, Mott insulators...). By contrast, the main limitation of our approach is the fact that it is based on a given Ansatz, which may contain a relevant bias. 

The second one is the Monte Carlo sampling: Computing expectation values of observables over the variational estimate of the ground-state wave function requires a numerical Monte Carlo sampling since the wave function is correlated. Moreover, the variational state contains many parameters to be optimized at the same time. This is a highly non-trivial and computationally demanding task that is made possible thanks to the Stochastic Reconfiguration method, which allows us to find the optimal set for hundreds of independent parameters~\cite{Sorella2005}. Once the optimal variational state is obtained, the presence of, for instance, spin order, charge order, and superconductivity can be detected with the computation of appropriate correlation functions. These physical observables are computed on finite lattice sizes with periodic boundary conditions and might require size scaling to the thermodynamic limit.

In Sec.~\ref{sec:model}, we briefly introduce the single-orbital Hubbard model and its multi-orbital extension. In this lecture, we will focus only on electronic models, even though the Hubbard model can also be extended to bosonic particles. In Sec.~\ref{sec:VMC}, we introduce the general scheme under which variational Monte Carlo works. Then, in  Sec.~\ref{sec:wf}, we present the variational wave functions that can properly describe superconductive and magnetic states in single-orbital and multi-orbital Hubbard models. We also detail how to perform Monte Carlo calculations for the given wave functions. In Sec.~\ref{sec:SR}, we describe how to optimize a variational wave function with the stochastic reconfiguration method. In Sec.~\ref{sec:results}, we present a few results for correlation functions that can be computed in variational Monte Carlo. Finally, in Sec~\ref{sec:final}, we draw the conclusions.

\section{The Hubbard model}\label{sec:model}

\subsection{The single-orbital Hubbard model}

\index{Hubbard model!single-orbital} In this section, we introduce the single-orbital Hubbard model on the square lattice, a prototype model for studying the role of electronic  correlations~\cite{Leblanc2015,qin2022}. In particular, this model is the simplest model able to capture the essential characteristics of the cuprates' phase diagram. This Hamiltonian comprises an interaction term $\mathcal{H}_{int}$ that contains the on-site electron-electron repulsion $U$ and a kinetic part $\mathcal{H}_{0}$ that includes the nearest neighbor hopping integral $t$, and the next-nearest neighbor hopping integral $t'$. This last term is directly connected with the critical temperature for superconductivity~\cite{pavarini2001}. The Hamiltonian is defined as:
\begin{equation}\label{eq:Hubbard_single}
     \mathcal{H}= \mathcal{H}_{0}+\mathcal{H}_{int}=-t\sum_{\langle R,R^\prime\rangle,\sigma}c^\dag_{R\sigma}c^{\dagga}_{R^\prime\sigma}-t^\prime\sum_{\langle\langle R,R^\prime\rangle\rangle,\sigma}c^\dag_{R\sigma}c^{\dagga}_{R^\prime\sigma} + \text{H.c} + U\sum_R n_{R\uparrow}n_{R\downarrow},
 \end{equation}
where $c^\dag_{R\sigma}$ and $c^{\dagga}_{R\sigma}$ denote the creation and annihilation operators of an electron with spin $\sigma$ on site $R$, respectively, while $n_{R\sigma}=c^\dag_{R\sigma}c^{\dagga}_{R\sigma}$ is the electron density per spin  $\sigma$ on site $R$. The symbols $\langle \cdots \rangle$ and $\langle \langle \cdots \rangle\rangle)$ represent the summation over nearest- and next-nearest-neighbor sites, respectively. By calling $N_e$ the total number of electrons and $L$ the total number of sites, the electron density is $n=N_e/L$. At half filling, i.e., when $n=1$, the model interpolates between a metal at $U=0$ and an insulator driven by electronic correlations, the so-called Mott insulator, at $t=0$. A metal-insulator transition will occur at a critical value of $U/t$ that depends on the lattice geometry.

Despite its simplicity, this model cannot be exactly solved, except in a few limiting cases, like in one spatial dimension with nearest-neighbor hopping, where the so-called Bethe Ansatz provides us with exact results~\cite{lieb1968}, and in the half-filled case for any value of $U/t$ at $t'=0$ on the square lattice, where the auxiliary-field Monte Carlo approach becomes numerically exact, with no sign problem~\cite{zhang1995}. In all other cases, we need to rely on approximate numerical methods. 

\subsection{The multi-orbital Hubbard model}

\index{Hubbard model!multi-orbital} When extending the Hubbard model from a single relevant orbital to a multi-orbital scheme, we need to include new interaction terms. In addition to the intra-orbital on-site Coulomb interaction $U$, we shall also include the inter-orbital on-site Coulomb repulsion $U'$ and the Hund's coupling $J$. The latter term tends to maximize the total spin on a single site, as a consequence of the intra-atomic exchange energy. The interaction term of the Hubbard Hamiltonian assumes then a more complicated form, also known as the Kanamori Hamiltonian~\cite{kanamori1963,georges2013}:
\begin{equation}\label{eq:HubbardKanamori}
\begin{split}
\mathcal{H}_{int} = & U \sum_{R} \sum_{\alpha} n_{R,\alpha,\uparrow} n_{R,\alpha,\downarrow} 
+U^\prime \sum_R \sum_{\alpha \neq \beta} n_{R,\alpha,\uparrow} n_{R,\beta,\downarrow} +(U^\prime-J) \sum_R \sum_{\alpha > \beta} \sum_{\sigma}  n_{R,\alpha,\sigma} n_{R,\beta,\sigma} \\
&-J \sum_R \sum_{\alpha \neq \beta} c^\dag_{R,\alpha,\uparrow} c^{\dagga}_{R,\alpha,\downarrow} c^\dag_{R,\beta,\downarrow} c^{\dagga}_{R,\beta,\uparrow} +J \sum_R \sum_{\alpha \neq \beta} c^\dag_{R,\alpha,\uparrow} c^\dag_{R,\alpha,\downarrow} c^{\dagga}_{R,\beta,\downarrow} c^{\dagga}_{R,\beta,\uparrow},
\end{split}
\end{equation}
where $\alpha$ and $\beta$ are orbital indices, $c^\dag_{R,\alpha,\sigma}$ ($c^{\dagga}_{R,\alpha,\sigma}$) is the fermionic operator that creates (annihilates) an electron in orbital $\alpha$, site $R$ and spin $\sigma$, and $n_{R,\alpha,\sigma}=c^\dag_{R,\alpha,\sigma} c_{R,\alpha,\sigma}$ is the density operator on site $R$, orbital $\alpha$ and spin $\sigma$. 

The first three terms refer to the on-site density-density interaction between electrons with opposite spins in the same orbital ($U$), opposite spins in different orbitals ($U'$), and parallel spins in different orbitals ($U' - J < U'$). In the latter term, the interaction constant is lower because of the spin alignment effect. The fourth is the spin-flip process ($J$), and the last term describes pair hopping, i.e., an up-down spin couple of electrons that moves from one orbital to another one ($J$). The preservation of rotational symmetry is guaranteed for $U' = U - 2J$, which will be used throughout the discussion.

The kinetic part of the multi-orbital Hubbard model depends on the specific hopping scheme that characterizes the model under investigation. Here, we consider iron-based superconductors as an example. In this case, a three Fe orbital ($d_{xz}$, $d_{yz}$, and $d_{xy}$) kinetic Hamiltonian ${\cal H}_{0}$~\cite{daghofer2010,fanfarillo2020} is able to reproduce qualitatively the generic shape and the orbital content of the Fermi surfaces of iron-based superconductors, while remaining sufficiently light to allow for accurate numerical simulations. The tight-binding Hamiltonian is then defined as:
\begin{equation}\label{eq:TB}
\mathcal{H}_{0} = \sum_k \sum_{\alpha,\beta} \sum_{\sigma} c^\dag_{k,\alpha,\sigma} T_{\alpha,\beta}(k) c^{\dagga}_{k,\beta,\sigma},
\end{equation}
where $\alpha$ and $\beta$ are orbital indices ($1=xz$, $2=yz$, and $3=xy$), while $c^\dagger_{k,\alpha,\sigma}$ ($c^{\dagga}_{k,\alpha,\sigma}$) is the fermionic 
operator that creates (annihilates) an electron in orbital $\alpha$, with momentum $k$ and spin $\sigma$. The intra- and inter-orbital hoppings are given as: 
\begin{eqnarray}
&&T_{1,1}(k) = 2t_2 \cos{k_x} + 2t_1 \cos{k_y} + 4t_3 \cos{k_x}\cos{k_y}, \nonumber \\
&&T_{2,2}(k) = 2t_1 \cos{k_x} + 2t_2 \cos{k_y} + 4t_3 \cos{k_x}\cos{k_y}, \nonumber \\
&&T_{3,3}(k) = 2t_5 (\cos{k_x} + \cos{k_y}) + 4t_6 \cos{k_x}\cos{k_y} + \epsilon_{xy},  \nonumber \\
&&T_{1,2} = T^{*}_{2,1} = 4t_4 \sin{k_x} \sin{k_y}, \nonumber \\
&&T_{1,3} = T^{*}_{3,1} = 2it_7 \sin{k_x} + 4it_8 \sin{k_x} \cos{k_y}, \nonumber \\
&&T_{2,3} = T^{*}_{3,2} = 2it_7 \sin{k_y} + 4it_8 \sin{k_y} \cos{k_x}. \nonumber
\end{eqnarray}
The hopping parameters (in units of eV) are set as: $t_1 = 0.02$, $t_2 = 0.06$, $t_3 = 0.03$, $t_4 = -0.01$, $t_5 = 0.1$, $t_6 = 0.15$, $t_7 = -0.1$, 
$t_8 = -t_7/2$~\cite{fanfarillo2020}. The two orbitals $d_{xz}$ and $d_{yz}$ are degenerate, while the orbital $d_{xy}$ has a crystal field $\epsilon_{xy} = 0.2$. 

\section{Variational Monte Carlo}\label{sec:VMC}

\index{variational Monte Carlo} Let us now introduce the general scheme under which variational Monte Carlo is defined. Let us first fix a complete orthonormalized basis in the Hilbert space $\{|x\rangle\}$:

\begin{equation}
 \sum_x |x \rangle \langle x |=\mathbb{I}.
\end{equation}

In order to clarify this statement with an example, an element of the basis for the single-orbital Hubbard model is a vector that contains the electronic occupation of each lattice site. 

Given the aforementioned basis, each quantum state can be written as:

\begin{equation}
 |\Psi\rangle=\sum_x |x\rangle \langle x |\Psi\rangle=\sum_x \Psi(x)|x\rangle.
\end{equation}

In this way, the expectation value of a generic operator $\mathcal{O}$ over a given variational state can be written as:

\begin{equation}\label{eq:expectation}
\langle\mathcal{O}\rangle=\frac{\langle \Psi|\mathcal{O}|\Psi\rangle}{\langle \Psi|\Psi\rangle}=\frac{\sum_x \langle\Psi|x\rangle\langle x|\mathcal{O}|\Psi\rangle}{\sum_{x'} \langle \Psi|x'\rangle \langle x' |\Psi\rangle}.
\end{equation}

The main problem in evaluating this expectation value lies in the number of configurations in the sum, which becomes exponentially large with the number of sites and particles. Therefore, for large sizes, it is impossible to perform an exact enumeration of the configurations to compute $\mathcal{O}$ exactly. However, Eq.~(\ref{eq:expectation}) can be recast into a form that can be easily treated by standard Monte Carlo methods:

 \begin{equation}
\langle \mathcal{O} \rangle =
\frac{\sum_x |\langle \Psi|x\rangle|^2 \frac{\langle x|\mathcal{O}|\Psi\rangle}{\langle x|\Psi\rangle}}{\sum_{x'}|\Psi(x')|^2}=
\frac{\sum_x |\Psi(x)|^2 O_L(x)}{\sum_{x'} |\Psi(x')|^2},
\end{equation}
where we have defined the local estimator:
\begin{equation}\label{eq:local}
O_L(x) = \frac{\langle x | \mathcal{O} | \Psi \rangle}{\langle x | \Psi \rangle}=\sum_{x'} \langle x|\mathcal{O}|x'\rangle \frac{\langle x'|\Psi\rangle}{\langle x|\Psi\rangle}.
\end{equation}
At first sight, the computation of the local estimator looks like a hard task, since it requires a summation over all the states of the many-body Hilbert space; however, when considering operators like the Hamiltonian, only a few terms contribute to the sum, thanks to the locality of the Hubbard Hamiltonian. Indeed, if we use the local basis, whose elements are the electronic configurations in the lattice, an element $|x\rangle$ is connected only to a few other configurations that differ in the hopping of one electron from a given site to one of its neighbors; then, the maximum number of such processes is equal to the number of sites $L$ times the number of bonds times the spin degeneracy. Therefore, the computation of the local estimator requires only a limited number of operations, usually proportional
to the number of sites or particles in the system.

The quantity
\begin{equation}\label{eq:probability}
P(x) = \frac{|\Psi(x)|^2}{\sum_{x'} |\Psi(x')|^2}
\end{equation}
can be interpreted as a probability distribution, since it is non-negative and is normalized. This property guarantees that variational Monte Carlo is not affected by the sign problem that is common to other Monte Carlo approaches.

Thus, the problem of computing a quantum average of the operator $\mathcal{O}$ can be rewritten as the average of the random variable $O_L(x)$, see Eq.~(\ref{eq:local}), over the probability distribution $P(x)$, see Eq.~(\ref{eq:probability}). In particular, if we consider the expectation value of the Hamiltonian, the local estimator is named local energy and is defined by:

\begin{equation}
e_L(x) = \frac{\langle x | \mathcal{H} | \Psi \rangle}{\langle x | \Psi \rangle}.
\end{equation}

It is then possible to define a stochastic algorithm (e.g., a Markov process) in which a sequence of electronic configurations $\{|x_n\rangle\}$ is generated. One can use, for example, the Metropolis algorithm, which is a customary algorithm in Monte Carlo simulations, for generating states according to the given probability distribution. With the Metropolis algorithm, after an initial equilibration time, configurations will be distributed according to the desired probability $P(x)$. Then, the quantum expectation value $\langle \mathcal{O}\rangle$ is evaluated as the mean value of the random variable $O_L(x)$  over the visited configurations:

\begin{equation}
\langle O \rangle \approx \frac{1}{N} \sum_{n=1}^{N} O_L(x_n), 
\end{equation}
where $N$ is the total number of visited configurations, after equilibration. 

An important feature of VMC is the zero-variance property: if the variational state $|\Psi\rangle$ coincides with an eigenstate of the Hamiltonian with eigenvalue $E$, the local energy $e_L(x)$ is a constant:

\begin{equation}
e_L(x) = \frac{\langle x | \mathcal{H} | \Psi \rangle}{\langle x | \Psi \rangle}=E \frac{\langle x | \Psi \rangle}{\langle x | \Psi \rangle}= E.
\end{equation}
 
 Therefore, the variance of the random variable $e_L(x)$ is zero. Even if this is an extreme case, it teaches us that the variance of $e_L(x)$ decreases when the variational state approaches an exact eigenstate. This fact is important to reduce statistical fluctuations and improve the numerical efficiency. Notice that the zero-variance property is a feature that exists only for quantum expectation values, while it is absent in classical calculations, where observables have thermal fluctuations. 

\subsection{The Metropolis algorithm}

\index{Metropolis algorithm} In this lecture, we do not discuss the details of Markov chains and the conditions that guarantee that the configurations $|x_n\rangle$ generated along the Markov chain are distributed according to the probability distribution $P(x)$ of Eq.~(\ref{eq:probability}), for large values of $n$. More details can be found, for instance, in Ref.~\cite{Becca2017}. Instead, we just recall the Metropolis algorithm that defines a transition probability for generating a new electronic configuration $|x'\rangle$, starting from an old one $|x\rangle$. Once a move is proposed to generate the new configuration $|x'\rangle$ starting from the old configuration $|x\rangle$, the acceptance probability $A(x'|x)$ is defined as:
\begin{equation}\label{eq:acceptance}
    A(x'|x) = \min \Big\{1, \; \frac{T(x|x') P(x')}{T(x'|x) P(x)} \Big\},
\end{equation}
where $T(x'|x)$ is the trial probability for generating $|x'\rangle$ from $|x\rangle$. If the trial probability is symmetric ($T(x'|x)=T(x|x')$), Eq.~(\ref{eq:acceptance}) further simplifies to: 
\begin{equation}
    A(x'|x) = \min \Big\{1, \; \frac{P(x')}{ P(x)} \Big\},
\end{equation}
where 
\begin{equation}\label{eq:Metropolis}
\frac{P(x')}{ P(x)}=\frac{|\Psi(x')|^2}{|\Psi(x)|^2}, 
\end{equation}
thus making it unnecessary to know the normalization condition of the wave function.

Recalling that each configuration $|x\rangle$ is the electron occupation of each lattice site (with a fixed $z$ component of the spin), we set in the following a generic procedure to run a Metropolis Monte Carlo algorithm for fermions:

\begin{itemize}
    \item Generate a random number to identify the sites $R$ and $R'$ among which the electron (with fixed spin $\sigma$) will hop, defining the trial probability $T(x'|x)$. In strongly correlated configurations (i.e., large values of $U/t$ at half filling), an additional trial move can be proposed, as a spin flip between two electrons with opposite spin on singly occupied sites. 
    \item If the electron can perform the move (specifically, it does not have to violate the Pauli principle), the algorithm computes the acceptance ratio as $A(x'|x)$; otherwise, the move is rejected.
    \item Generate a random uniform number $\eta \in (0,1]$ to compare with the acceptance ratio. The new configuration $|x'\rangle$ is accepted for $A(x'|x) \geq \eta$; otherwise, the move is rejected.
    \item Once thermalization is reached, i.e., when configurations are generated according to the desired probability distribution, observables, such as the local energy $E_L(x)$, can be computed every $O(L)$ steps, where $L$ is the lattice size, to have uncorrelated configurations. As a rule of thumb, thermalization time is similar to correlation time, that is, the necessary time to run a simulation from a starting configuration $\ket{x}$ to obtain an independent configuration $\ket{x'}$.
\end{itemize}

\index{error bars} If the configurations are uncorrelated, the error bar associated with the variational estimate of an observable is given by:

\begin{equation}\label{eq:error}
    \sigma_{\langle \mathcal{O}\rangle} = \sqrt{\frac{1}{N(N-1)} \Big(  \sum_{i = 1}^N O_L^2(x_i) -  \frac{1}{N} \Big(\sum_{i = 1}^N O_L(x_i)\Big)^2 \Big) }.
\end{equation}

However, the usual acceptance probability for VMC in the Hubbard model is quite small, implying that many moves are rejected, and the electronic configurations stored during the Monte Carlo sampling could be correlated. Consequently, the error bar will be underestimated if we use Eq.~(\ref{eq:error}). A simple procedure used to solve the problem of successive correlated configurations, in the evaluation of the error bar, is the \textit{binning technique}, in which $L_b$ measures are grouped in a bin, in order to compute a partial average within the bin $b$:

\begin{equation}
    \bar{O_b} = \frac{1}{L_b} \sum_{i = (b - 1)L_b + 1}^{b L_b} O_L(x_i);
\end{equation}

here, $b = 1 \ldots N_{bin}$ such that $N = \sum_{b=1}^{N_{bin}} L_b$. Whenever $L_b$ is larger than the correlation time, the configurations are almost uncorrelated. Based on the binning technique, the variational energy and its error bar are given by:

\begin{gather}
    \langle \mathcal{O}\rangle = \frac{1}{N_{bin}} \sum_{b = 1}^{N_{bin}} \bar{O_b} \nonumber \\
    \sigma_{\langle \mathcal{O\rangle}} = \sqrt{\frac{1}{N_{bin}(N_{bin} - 1)} \sum_{b=1}^{N_{bin}} (\bar{O_b} -  \langle\mathcal{O}\rangle)^2  }.
\end{gather}

\section{Variational wave functions}\label{sec:wf}

\subsection{The Jastrow-Slater wave function for the single-orbital Hubbard model}

\index{wave function!single-orbital} The Jastrow-Slater wave function can be generically written as:

\begin{equation}
 |\Psi_{J}\rangle=\mathcal{J}|\Phi_0\rangle,
\end{equation}
where $\mathcal{J}$ is a Jastrow factor that embeds correlation in the wave function and $|\Phi_0\rangle$ is the ground state of a noninteracting quadratic auxiliary Hamiltonian $\mathcal{H}_{aux}$, according to the definition:

\begin{equation}
    |\Phi_0\rangle=\prod_{k=1}^{N_e} \phi_k^\dagger|0\rangle,
\end{equation}
where $N_e$ is the total number of electrons and $\phi_k$ are the eigenstates of $\mathcal{H}_{aux}$. Indeed, we can easily diagonalize any Hamiltonian that can be written in a simple quadratic form as:

\begin{equation}
\mathcal{H}_{aux}=\mathbf{d}^{\dag} \mathbf{T}\mathbf{d},
\end{equation}
where $\mathbf{d}^\dag=(\mathbf{d}^\dag_{\uparrow},\mathbf{d}^\dag_{\downarrow})$ is a vector with $2L$ components $(\mathbf{d}^\dag_{\sigma}=d^\dag_{1,\sigma},\dots,d^\dag_{L,\sigma})$, $\mathbf{T}$ is a $2L\times 2L$ matrix, and $L$ is the number of lattice sites. By introducing a unitary matrix $\mathbf{U}$ (that preserves the anticommutation rules of fermions), we can diagonalize $\mathcal{H}_{aux}$:
\begin{equation}
\mathcal{H}_{aux}=\mathbf{d}^{\dag}\mathbf{U}\mathbf{U}^\dag \mathbf{T}\mathbf{U}\mathbf{U}^\dag\mathbf{d}=\mathbf{\Phi}^\dag \mathbf{E}\mathbf{\Phi}=\sum_k\epsilon_k\phi_k^\dag\phi^{\dagga}_k,
\end{equation}
where $\mathbf{E}$ is a diagonal matrix that contains the eigenvalues and $\mathbf{\Phi}=(\phi_1,\dots\phi_{2L})$ is defined in terms of the eigenvectors.

\index{auxiliary Hamiltonian} The choice of the noninteracting quadratic Hamiltonian is a careful aspect, since we need to include in it the relevant physical terms that we want to study in our model. For instance, if one would be interested in studying the emergence of superconductivity in the Hubbard model, when varying the electronic density, and its competition with antiferromagnetic N\'eel order, one could define the following auxiliary Hamiltonian:

\begin{equation}
    \mathcal{H}_{aux}=\mathcal{H}_{0}+\mathcal{H}_{AF}+\mathcal{H}_{BCS}.
\end{equation}
The first term consists of the \textit{kinetic energy} of the electrons in the Hubbard model:
 \begin{equation}
     \mathcal{H}_{0}= -t\sum_{\langle R,R^\prime\rangle,\sigma}c^\dag_{R\sigma}c^{\dagga}_{R^\prime\sigma}-\Tilde{t^\prime}\sum_{\langle\langle R,R^\prime\rangle\rangle,\sigma}c^\dag_{R\sigma}c^{\dagga}_{R^\prime\sigma} + \text{H.c}.
 \end{equation}
The second term includes antiferromagnetism by \textit{standard N\'eel order}
\begin{equation}
    \mathcal{H}_{AF}=\Delta_{AF} \sum_{R}(-1)^{x+y}\left(c^\dag_{R\uparrow}c^{\dagga}_{R\uparrow}-c^\dag_{R\downarrow}c^{\dagga}_{R\downarrow}\right),
\end{equation}
while the last one introduces the \textit{BCS electron pairing}:
\begin{equation}
    \mathcal{H}_{BCS}=\sum_{R,\eta=x,y}\Delta_{\eta}\left(c^\dag_{R,\uparrow}c^\dag_{R+\eta,\downarrow}-c^\dag_{R,\downarrow}c^\dag_{R+\eta,\uparrow}\right)+\text{H.c.}-\mu\sum_{R,\sigma}c^\dag_{R\sigma}c^{\dagga}_{R\sigma},
\end{equation}
where $\mu$ is a chemical potential and $\Delta_x=-\Delta_y$ represents the usual $d$-wave symmetry that is realized in the single-band Hubbard model. Further developments can be included to describe the striped modulations of charge and spin that are a common feature in the ground state of the single-orbital Hubbard model on the square lattice~\cite{zheng2017,darmawan2018,marino2022}. The parameters $\tilde{t'},\Delta_{AF},\Delta{\eta},\mu$ are variational parameters that can be optimized, as described in Sec .~\ref{sec:SR}, while we keep $t=1$ to fix the energy scale. 

The reader could now wonder how a BCS state can be written as a Slater determinant. This is indeed possible, via a convenient particle-hole canonical transformation of the annihilation and creation operators for down spins, i.e.:

\begin{equation}
    d^{\dagger}_{R,\uparrow} \equiv c^{\dagger}_{R,\uparrow} \quad
    d_{R,\downarrow} \equiv c^{\dagger}_{R,\downarrow}.
\end{equation}
The products of creation and annihilation operators that build up the auxiliary Hamiltonian are then rewritten as:
\begin{equation}  
\begin{split}
&c^\dag_{R\uparrow}c^{\dagga}_{R^\prime\uparrow}=d^\dag_{R\uparrow}d^{\dagga}_{R^\prime\uparrow} \quad c^\dag_{R\downarrow}c^{\dagga}_{R^\prime\downarrow}=-d^\dag_{R'\downarrow}d^{\dagga}_{R\downarrow} \\
&c^\dag_{R,\uparrow}c^\dag_{R+\eta,\downarrow}=c^\dag_{R,\uparrow}c^{\dagga}_{R+\eta,\downarrow} \quad c^\dag_{R,\downarrow}c^\dag_{R+\eta,\uparrow}=-c^\dag_{R+\eta,\uparrow}c^{\dagga}_{R,\downarrow}.
\end{split}
\end{equation}
In this way, the eigenstates of the BCS Hamiltonian can be expressed as “orbitals”, similarly to the ones that can be obtained by diagonalizing the kinetic term.

It should be noted that the particle-hole transformation changes the total number of electrons per spin species $N_{\sigma}^c$, according to the following relations:

\begin{gather}
    N_{\uparrow}^c = \sum_R c_{R,\uparrow}^\dag c^{\dagga}_{R,\uparrow} = \sum_R d_{R,\uparrow}^\dag d^{\dagga}_{R,\uparrow} = N_{\uparrow}^d \nonumber \\
    N_{\downarrow}^c = \sum_R c_{R,\downarrow}^\dag c^{\dagga}_{R,\downarrow} = L - \sum_R d_{R,\downarrow}^\dag d^{\dagga}_{R,\downarrow} = L - N_{\downarrow}^d,
\end{gather}
where $L$ is the total number of sites.

Whenever the total magnetization is zero, i.e., $N^c_{\uparrow}=N^c_{\downarrow}$, the total number of electrons after the particle-hole transformation, given by $ N_{\uparrow}^d + N_{\downarrow}^d$, is equal to $L$, that is, the system, after the particle-hole transformation, is always half-filled.

\index{Jastrow factor} The \emph{Jastrow factor} $\mathcal{J}$ is a long-range correlator that on the lattice takes the following form:

\begin{equation}\label{eq:Jastrow}
 \mathcal{J}=\exp\left[-\frac{1}{2} \sum_{R,R'}v_{R,R'} (n_R-n)(n_{R'}-n)\right],
\end{equation}
where $v_{R,R'}$ is a pseudo-potential for density-density correlations in the variational state, $n_R$ is the electronic density on site $R$, and $n$ is the average electronic density. For translationally invariant models, like the Hubbard Hamiltonian of Eq.~(\ref{eq:Hubbard_single}), $v_{R,R'}$ only depends on the relative distance of the two sites $|\mathbf{R}-\mathbf{R'}|$. The Jastrow pseudo-potential can be optimized for all possible distances, which are $O(L)$ in translationally invariant
systems, via the stochastic reconfiguration method of Sec.~\ref{sec:SR}. The role of the long-range tail of the Jastrow 
factor is to create a bound state between doubly occupied sites and empty sites, thus not allowing for conduction, but still allowing local density fluctuations. Indeed, it has been shown that the Jastrow term may turn a non-interacting metallic state $|\Phi_0\rangle$ at half filling into a Mott insulator, that is, an insulator driven by electronic correlation, even in the absence of magnetic order~\cite{capello2005}. We mention the existence of a deep connection between a Mott insulator at half-filling and a superconductor at finite doping in the Resonant Valence Bond (RVB) theory of high-temperature superconductivity proposed by Anderson~\cite{anderson1987}. He suggested that a superconducting phase may emerge when doping a Mott insulator with ``preformed'' singlet
pairs. In fact, the RVB state describes an insulator that is a liquid of spin singlets, which become mobile when the system is doped, thus forming actual superconducting pairs. 

\index{Gutzwiller factor} We remark that replacing the Jastrow factor with a much simpler term that penalizes the formation of double occupancies is not able to reproduce a Mott insulator unless $U=\infty$. This simplified interaction term, called the Gutzwiller factor, is defined as~\cite{gutzwiller1963}:

\begin{equation}
 |\Psi_G\rangle=P_G|\Phi_0\rangle \quad P_G=\exp\left[-\frac{g}{2}\sum_R (n_R-n)^2\right]
\end{equation}
and corresponds to the $R=R'$ case of the Jastrow factor. 
The effect of the Gutzwiller factor becomes clear once the variational state is expanded in a basis set whose elements $\{|x\rangle\}$ represent configurations with
electron occupations of each lattice site. Since the Gutzwiller factor is diagonal in this basis, we
have that:

\begin{equation}
\langle x|\Psi_G\rangle = P_G(x) \langle x|\Phi_0\rangle,
\end{equation}
where $P_G(x)\le 1$ is a number that depends on how many doubly-occupied sites are present in a given electronic configuration. Therefore, the amplitude of the non-interacting state $|\Phi_0\rangle$ is renormalized by $P_G (x)$. At half filling, density excitations are represented
by doublons (doubly-occupied sites) and holons (empty sites). In the absence of electronic correlation, these objects are free to move and are then responsible
for conductivity (for example, in the fermionic model, a doublon is negatively charged with respect to the average background, while the holon is
positively charged). The effect of the Gutzwiller factor is to penalize the formation of such excitations; however, once created, doublons and holons are no longer correlated, thus being free to move independently. Only when the energetic penalty is infinite, an insulator is obtained. On the contrary, the long-range tail of the Jastrow factor allows to correlate holons and doublons and keeps them bound together. This fact can be understood via a classical mapping to a model of charged particles, where holons and doublons represent positive and negative charges~\cite{Becca2017}. 

\subsection{Computing ratios for the Jastrow-Slater wave function}

\index{variational Monte Carlo} Both in the Metropolis algorithm of Eq.~(\ref{eq:Metropolis}) and in the evaluation of the local observables of Eq.~(\ref{eq:local}), we need to compute ratios of amplitudes of the variational state with respect to a generic element of the basis set, i.e., $\langle x'|\Psi\rangle/\langle x|\Psi\rangle$.

The advantage of considering Jastrow-Slater wave functions in the variational Monte Carlo technique is that the calculations can be efficient and fast. Indeed, since the Jastrow factor is diagonal in the
chosen basis, we have that:
\begin{equation}
    \langle x|\Psi\rangle=\mathcal{J}(x)\langle x|\Phi_0\rangle,
\end{equation}
where $\mathcal{J}(x)$ is the value of the Jastrow operator computed for the configuration $|x\rangle$, i.e., $\mathcal{J}|x\rangle=\mathcal{J}(x)|x\rangle$. Therefore, given the electronic configuration, $\mathcal{J}(x)$ is a number that can be evaluated in $O(L^2)$ operations. Moreover, $\langle x|\Phi_0\rangle$ can be evaluated in $O(L^3)$ operations since it is a Slater determinant.

Let us now briefly discuss how to compute ratios for the two cases. First, let us focus on the ratio of Jastrow factors $\mathcal{J}(x')/\mathcal{J}(x)$ and show that it requires only $O(L)$ operations. We focus on a simplified form for the Jastrow factor:
\begin{equation}
   \mathcal{J}=\exp\left[-\frac{1}{2} \sum_{R,R'}v_{R,R'} n_Rn_{R'}\right],
\end{equation}
which is equivalent to the one of Eq.~(\ref{eq:Jastrow}) when the particle number is conserved (apart from a multiplicative factor).

When $|x'\rangle$ is close to $|x\rangle$, i.e.: 
\begin{equation} |x\rangle\equiv|n_1,\dots,n_k,\dots,n_l,\dots,n_L\rangle \qquad |x'\rangle\equiv|n_1,\dots,n_{k+1},\dots,n_{l-1},\dots,n_L\rangle,
\end{equation}

it is possible to define a fast computation:
 \begin{equation}\label{eq:Jratio}
  \frac{{\cal J}(x')}{{\cal J}(x)}=\frac{{\cal J}(n_1,\dots,n_{k+1},\dots,n_{l-1},\dots,n_L)}{{\cal J}(n_1,\dots,n_k,\dots,n_l,\dots,n_L)}=\frac{\exp(-\sum_R v_{R,k}n_R)}{\exp(-\sum_R v_{R,l}n_R)}e^{v_{k,l}-v_{k,k}},
 \end{equation}
where we used the fact that the pseudo-potential is symmetric and translationally invariant. The term $e^{v_{k,l}-v_{k,k}}$ in Eq.~(\ref{eq:Jratio}) does not depend upon the electronic configuration and
can be computed at the beginning of the simulation, once for all. Instead,
the ratio of the exponentials depends on the fermionic configuration, which is sampled
along the Markov chain. However, the computation of the ratio can be done in $O(1)$ operations once we compute and store a vector of dimension $L$ that depends on $|x\rangle$, defined as:

\begin{equation}\label{Tvector}
    T_{\textrm{Jastrow}}(R')=\sum_Rv_{R,R'}n_R.
\end{equation}
Then, updating $T_{\textrm{Jastrow}}(R')$ to the new configuration $|x'\rangle$ costs $O(L)$:
 \begin{equation}
  n'_R=n_R+\delta_{k,R}-\delta_{l,R} \Rightarrow T'_{\textrm{Jastrow}}(R')=T_{\textrm{Jastrow}}(R')+v_{k,R'}-v_{l,R'}.
 \end{equation}
Therefore, we can compute from scratch $T_{\textrm{Jastrow}}(R')$ at the beginning of the Markov chain for all the sites $R' = 1,\dots, L$ and then update it every time a new configuration is accepted along the Markov chain. From time to time, it is recommended to recompute Eq.~(\ref{Tvector})
from scratch, since the rounding error of the update can accumulate and give rise to numerical errors.

Let us now move to the $\langle x|\Phi_0\rangle$ term, and show that it indeed corresponds to a Slater determinant that can be computed in $O(L^3)$ operations with standard linear algebra routines. First of all, let us contract the site and spin indices of the auxiliary quadratic Hamiltonian $\mathcal{H}_{aux}$ into a single index $I$ from 1 to $2L$: $c_{R,\uparrow}\equiv d_{R}\quad c_{R,\downarrow}\equiv d_{R+L}$. Then, we recall that ${\cal H}_{aux}$ can be diagonalized and its eigenstates are the single-particle orbitals $\phi^{\dagger}_{k}|0\rangle$:
  \begin{equation}
   {\cal H}_{aux}=\sum_{k}\epsilon_{k}\phi^{\dagger}_{k}\phi_{k},
  \end{equation}
 with $\epsilon_{k}$ the single-particle energies and $\phi_{k}^{\dagger}=\sum_I U_{I,k}d^{\dagger}_I$ (where $\mathbf{U}$ is the unitary matrix that diagonalizes ${\cal H}_{aux}$). Then, the many-particle state $|\Phi_0\rangle$ can be constructed by occupying the $N_e$ lowest single-particle orbitals, where $N_e$ is the number of electrons:
 \begin{equation}
|\Phi_0\rangle=\prod_{k=1}^{N_e}\phi_{k}^{\dagger}|0\rangle=\prod_{k=1}^{N_e}\left( \sum_I U_{I,k}d^{\dagger}_I\right)|0\rangle.
 \end{equation}
A generic electronic configuration for $N_e$ electrons in the lattice, visited along the Markov chain, can be written as:
\begin{equation}
 |x\rangle=d^{\dagger}_{R_1}\dots d^{\dagger}_{R_{N_e}}|0\rangle,
 \end{equation}
 where a generic $R_j$ assumes values
from 1 to $2L$: the positions of spin-up electrons coincide with the site number,
while the positions of spin-down electrons must be shifted by $L$. The overlap $\langle x|\Phi_0\rangle$ can then be written as: 
\begin{equation}\begin{split}
\langle x|\Phi_0\rangle&=\langle 0|d_{R_{N_e}}\dots d_{R_1}\left( \sum_I U_{I,1}d^{\dagger}_I\right)\dots \left(\sum_I U_{I,N_e}d^{\dagger}_I\right)|0\rangle= \\
&= \langle 0|d_{R_{N_e}}\dots d_{R_1}\left[\sum_p (-1)^p\prod_{k=1}^{N_e} U_{p\{R_j\},k}\right]  d^{\dagger}_{R_1}\dots d^{\dagger}_{R_{N_e}}|0\rangle.
\end{split}\end{equation}

The sum in $[\dots]$ is over all the permutations of the $\{R_j\}$ in $|x\rangle$ (with the sign from anticommutations), thus implying that  $\langle x|\Phi_0\rangle=\det\{U_{R_j,k}\}$. In practice, the diagonalization of the auxiliary Hamiltonian can be done at the beginning of the simulation, obtaining the matrix of the eigenvectors $\mathbf{U}$, of which we store only the $N_e$ columns corresponding to the occupied orbitals. Then, the overlap $\langle x|\Phi_0\rangle$ with the generic configuration $|x\rangle$ is given by the determinant of the matrix $\tilde{\mathbf{U}}$:
\begin{equation}
    \tilde{\mathbf{U}} =
\begin{pmatrix}
U_{R_1,1} & \cdots & U_{R_1,N_e} \\
\vdots & \ddots & \vdots \\
U_{R_l,1} & \cdots & U_{R_l,N_e} \\
\vdots & \ddots & \vdots \\
U_{R_{N_e},1} & \cdots & U_{R_{N_e},N_e}
\end{pmatrix},
\end{equation}
obtained by  taking only the rows corresponding to the occupied sites of the matrix $\mathbf{U}$. Now, if we consider a new electronic configuration $|x'\rangle$, obtained from $|x\rangle$ after the hopping of the $l$-th electron from position $R_l$ to position $R_{l'}$, i.e., $|x'\rangle=d^\dag_{R_{l'}}d^{\dagga}_{R_l}|x\rangle$, we obtain a new matrix $\tilde{\mathbf{U'}}$ that will differ from $\tilde{\mathbf{U}}$ for the $l$-th row:
\begin{equation}
    \tilde{\mathbf{U'}} =
\begin{pmatrix}
U_{R_1,1} & \cdots & U_{R_1,N_e} \\
\vdots & \ddots & \vdots \\
U_{R_l',1} & \cdots & U_{R_l',N_e} \\
\vdots & \ddots & \vdots \\
U_{R_{N_e},1} & \cdots & U_{R_{N_e},N_e}
\end{pmatrix}.
\end{equation}
Then, the ratio between two configurations that differ only by the hopping of a single electron is given by:
\begin{equation}
\frac{\langle x' | \Phi_0 \rangle}{\langle x | \Phi_0 \rangle}
=
\frac{\langle x | d^{\dagger}_{R_l} \, d_{R_l'} | \Phi_0 \rangle}{\langle x | \Phi_0 \rangle}
=
\frac{\det \tilde{\mathbf{U}}'}{\det \tilde{\mathbf{U}}} \,.
\end{equation}
It is then possible to show that this ratio of determinants can be computed in only $O(L^2)$ operations, thus speeding up the simulation with respect to a full calculation of a determinant from scratch~\cite{Becca2017}.

\subsection{The backflow terms}

\index{backflow} While the Jastrow-Slater wave function is able to capture Mott insulating and superconductive states, it has a strong intrinsic limitation: Electronic correlations are introduced only via a multiplicative factor, while the nodes of the variational state are the ones set by the uncorrelated state $|\Phi_0\rangle$. In order to overcome this limitation, we have introduced a novel variational state that redifines the single-particle orbitals of the uncorrelated state, taking into account the electronic repulsion, the so-called backflow correlations~\cite{tocchio2008,tocchio2011}. 

Let us consider the single-particle eigenstates $\phi_k(R,\sigma)$ of the auxiliary Hamiltonian $\mathcal{H}_{aux}$, written in the basis of lattice sites and $z$ component of the spins. Backflow correlations modify them,  according to the electronic configuration on the 
lattice:
\begin{equation}\label{eq:backflow}
\phi_{k}^{b}(R,\sigma)  \equiv  \tilde{\epsilon}
\phi_{k}(R,\sigma)+ \eta_1 \sum_{R' \textrm{n.n.} R}
D_RH_{R'} \phi_{k}(R',\sigma) 
+  \eta_2 \sum_{R' \textrm{n.n.n.} R}
D_R H_{R'} \phi_{k}(R',\sigma),
\end{equation}
where $\tilde{\epsilon}=\epsilon$ if the site $R$ is doubly occupied and is 
surrounded by at least one empty site, while $\tilde{\epsilon}=1$ in all the
other cases. Here $\epsilon, \eta_l,(l=1,2)$ are variational parameters to be 
optimized, according to the method defined in Sec.~\ref{sec:SR}. The operators $D_R=n_{R,\uparrow}n_{R,\downarrow}$ and 
$H_{R}=h_{R,\uparrow}h_{R,\downarrow}$, with $h_{R,\sigma}=1-n_{R,\sigma}$, are non-zero only if the site $R$ is doubly occupied or empty, respectively. Thus, they are the operators that allow us to include many-body correlations inside the single-particle orbitals. Moreover, the shorthand notations n.n. and n.n.n. indicate nearest- and next-nearest-neighbor sites, respectively. In this way, already the determinant
part of the wave function includes correlation effects, strongly improving the accuracy of the many-body state. 
This is a substantial improvement with respect to Jastrow factors, where electron-electron correlation is included only via a multiplicative term.

\begin{figure}[t!]
\centering
 \includegraphics[width=0.49\textwidth]{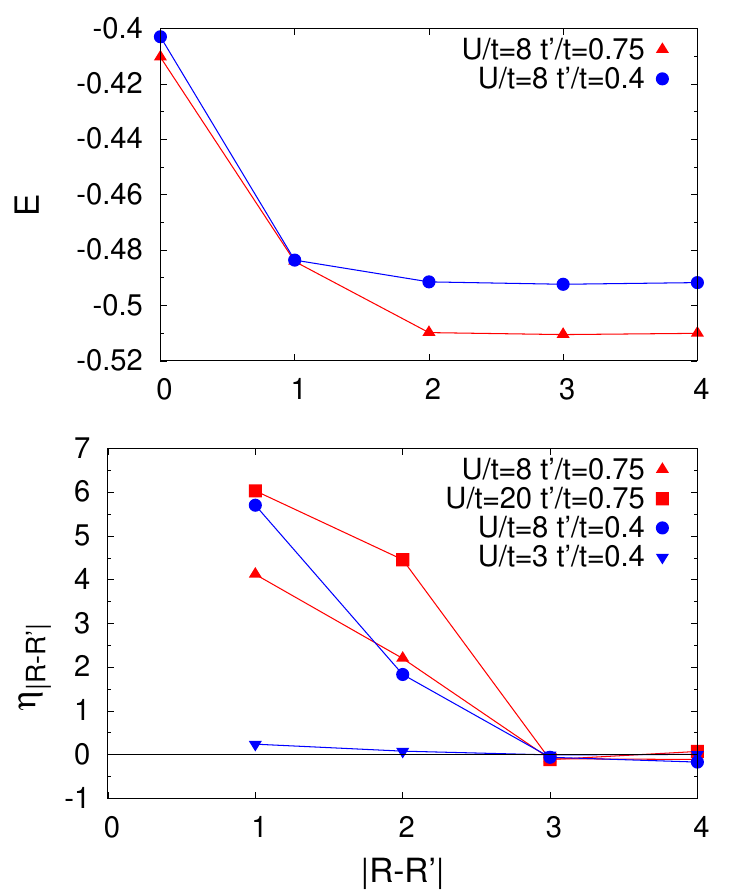}
 \includegraphics[width=0.49\textwidth]{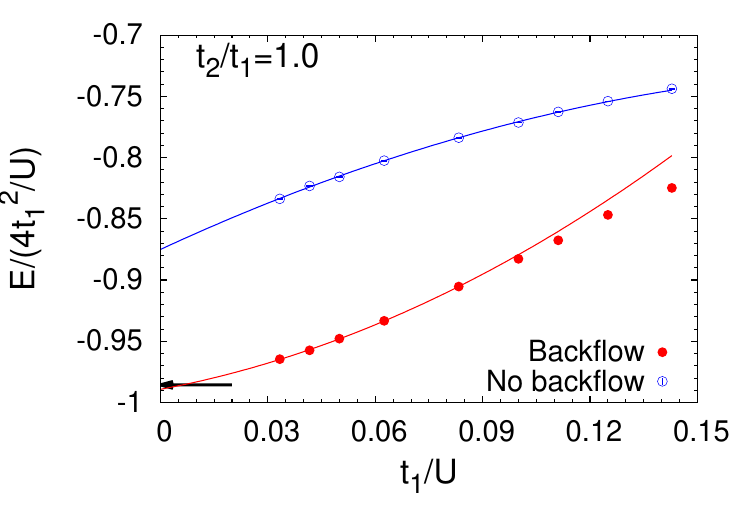}
 \caption{Upper left panel: Ground-state energy as
a function of the range of backflow parameters on the 2D square lattice for $t'/t=0.4$ and $0.75$ at $U/t = 8$. A range equal to zero corresponds to no backflow correlations in the wave function. Lower left panel: Optimized backflow parameters
$\eta_{|\mathbf{R}-\mathbf{R'}|}$ as a function of the distance $|\mathbf{R}-\mathbf{R'}|$ between lattice sites. Results for $U/t = 3$ and $t'/t=0.4$ correspond to a metallic state, the others to Mott insulating states. Data are shown for a lattice with $L=98$ sites~\cite{tocchio2011}. Right panel: Energy (in units of $4t_1^2/U)$, as a function of $t_1/U$, for the 1D Hubbard model with $t_2/t_1 = 1$. Filled (empty) circles denote the results with (without) backflow correlations. The DMRG energy of the corresponding Heisenberg model with $J_2/J_1 = 1$ is shown by an arrow~\cite{chitra1995}. Data are presented for an $L = 120$ lattice size~\cite{tocchio2011}.}
 \label{fig:backflowa}
\end{figure}

The terms in Eq.~(\ref{eq:backflow}) are the dominant contributions to the backflow operator, having a qualitative influence on the properties of the 
correlated wave functions. In addition, we can also take into account further terms that are useful in the intermediate-coupling regime and correspond to all possible hopping processes:
\begin{eqnarray}\label{eq:backflow2}
&& \phi_{k}^{b}(R,\sigma) \equiv \tilde{\epsilon}
\phi_{k}(R,\sigma)+ \eta_1 \sum_{R' \textrm{n.n.} R}
D_R H_{R'} \phi_{k}(R',\sigma) + \nonumber \\
&& \eta_2 \sum_{R' \textrm{n.n.n.} R}
D_R H_{R'} \phi_{k}(R',\sigma) + \nonumber \\
&& \beta_1 \sum_{R' \textrm{n.n.} R}
n_{R,\sigma}h_{R,-\sigma}n_{R',-\sigma}h_{R',\sigma}
\phi_{k}(R',\sigma)  + \nonumber \\
&& \beta_2 \sum_{R' \textrm{n.n.n.} R}
n_{R,\sigma}h_{R,-\sigma}n_{R',-\sigma}h_{R',\sigma}
\phi_{k}(R',\sigma) + \nonumber \\ 
&& \gamma_1 \sum_{R' \textrm{n.n.} R}
\left( D_R n_{R',-\sigma}h_{R',\sigma} + n_{R,\sigma} h_{R,-\sigma}H_{R'}\right)
\phi_{k}(R',\sigma) + \nonumber \\
&& \gamma_2 \sum_{R' \textrm{n.n.n.} R}
\left( D_R n_{R',-\sigma}h_{R',\sigma} + n_{R,\sigma} h_{R,-\sigma}H_{R'}\right)
\phi_{k}(R',\sigma),
\end{eqnarray}
where, in addition to $\epsilon$ and $\eta_l$, $\beta_l,\gamma_l (l=1,2)$ are 
variational parameters to be optimized. We mention that the computation of $\langle x|\Phi_0\rangle$ and of the determinant ratios introduced in the previous subsection can be extended to the presence of backflow correlations with a similar degree of complexity~\cite{Becca2017}.

Let us now present a few numerical results that show the relevance of backflow correlations in describing electronic correlations. We focus on the half-filled case, where the effect of electronic correlations is maximal, and we consider the simpler single-orbital Hubbard model. Moreover, we work in the so-called ``nonmagnetic sector", that is, neglecting any possible magnetic order in the variational wave function. In this way, we focus on the appearance of Mott insulating states that are driven solely by electronic correlations. 

\begin{figure}[t!]
\centering
 \includegraphics[width=0.9\textwidth]{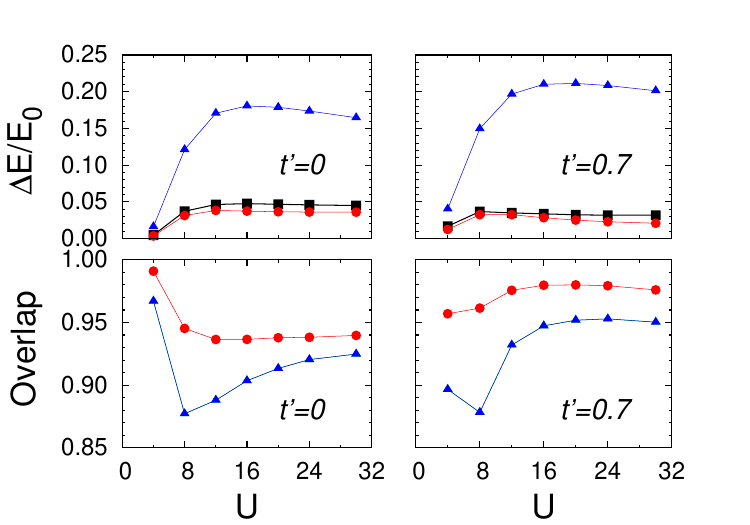}
 \caption{Results for 18 electrons on 18 sites as a function of $U/t$. Upper panels: Accuracy of energy $\Delta E = (E_0-E_v)$, $E_v$ and $E_0$ being the variational and the exact values, respectively. Lower panels: Overlap between the exact ground state and the variational wave functions. The variational state without backflow is denoted by blue triangles, the variational state with backflow correlations is denoted by red circles~\cite{tocchio2008}.}
 \label{fig:backflowb}
\end{figure}

First, let us discuss the range of backflow correlations. We present the results in 
Fig.~\ref{fig:backflowa}, left panels, for a 2D square lattice with both nearest neighbor hopping $t$ and next-nearest neighbor hopping $t'$. We report that backflow parameters are irrelevant for all distances larger than the ones connected by the hopping amplitude. Most importantly, a remarkable gain in energy is obtained only by a suitable optimization of the backflow parameters up to second neighbors, while poor results are obtained when there is no backflow (or its range is smaller than the one of the hopping). Moreover, in Fig.~\ref{fig:backflowa}, right panel, we present the effect of backflow correlations in the large-$U$ limit. We consider a one-dimensional Hubbard model, which is the same as the one introduced in Eq.~(\ref{eq:Hubbard_single}), just denoting the nearest-neighbor hopping with $t_1$ and the next-nearest-neighbor hopping with $t_2$. Only the presence of backflow correlations in the wave function allows for a proper extrapolation to the infinite-$U$ limit of the Hubbard model, that is, the Heisenberg model. Here, the energy of the Heisenberg model is provided by density-matrix renormalization group (DMRG) calculations, which are numerically exact for 1D models. We also compare the variational results with the exact ones on the 18-site cluster. In Fig.~\ref{fig:backflowb}, we show the accuracy of the variational state (with and without backflow correlations) and the overlap with the exact ground state for two hopping values, i.e., $t'/t = 0$ and $t'/t=0.7$. The backflow term is able to highly improve the accuracy both for weak and strong couplings. Moreover, the overlap between the exact ground state and the backflow state remains very high, even for large values of $U/t$, and the improvement with respect to the state without backflow is crucial, especially when a finite value of $t'$ is included.

These results suggest that backflow correlations are indeed necessary to properly describe electronic correlations. In particular, backflow correlations turned out to be particularly useful in the presence of frustration~\cite{tocchio2008,tocchio2013}. We do not discuss frustration in this lecture, but let us just say that it is related to the presence of competing hopping parameters in Mott insulators, like the case of square lattices with both nearest neighbor and next-nearest neighbor hopping parameters. The presence of frustration may destroy magnetic orders and lead to the formation of spin liquid phases, i.e., Mott insulating states without long-range magnetic order. Moreover, the backflow wave function has been the starting point for constructing a new class of wave functions, named neural network backflow, which uses a neural network to learn the optimal transformation via variational Monte Carlo calculations~\cite{luo2019}.

\subsection{The Jastrow-Slater wave function for the multi-orbital Hubbard model}

\index{wave function!multi-orbital} The Jastrow-Slater wave function for a multi-orbital Hubbard model can be generalized in the following form~\cite{marino2025,marino2025b}:

\begin{equation} |\Psi\rangle=\mathcal{J}_U\mathcal{J}_{U'}\mathcal{J}_J|\Phi_0\rangle.
\end{equation} 

The Jastrow factor has been generalized as:

\begin{gather}
   \mathcal{J}_U = \exp \Big(-\frac{1}{2}  \sum_{\alpha} \sum_{R,R'} v_{R,R'}^{\alpha,\alpha} n_{\alpha,R} n_{\alpha,R'}\Big) \nonumber \\
   \mathcal{J}_{U'} = \exp \Big(-\frac{1}{2}  \sum_{\alpha \neq \beta} \sum_{R,R'} v_{R,R'}^{\alpha,\beta} n_{\alpha,R} n_{\beta,R'}\Big).
\end{gather}

When $\alpha  = \beta$, the Jastrow factor is intra-orbital with a pseudopotential $v_{R,R'}^{\alpha,\alpha}$, whereas for $\alpha \neq \beta$ the electronic correlations are inter-orbital and its pseudopotential is written as $v_{R,R'}^{\alpha,\beta}$. The Jastrow factors mimic the Coulomb interaction present in the Hubbard-Kanamori Hamiltonian of Eq.~(\ref{eq:HubbardKanamori}). Specifically, $\mathcal{J}_U$ and $\mathcal{J}_U'$ take into account the presence of the intra-orbital repulsion $U$ and the inter-orbital repulsion $U'$, respectively. In order to take into account the presence of the  Hund's coupling in the wave function, we define a spin-spin Jastrow factor that favors configurations where electron spins in different orbitals (and same site) are aligned:

\begin{equation}
    \mathcal{J}_{J} = \exp \Big(- \frac{1}{2} \sum_{\alpha \neq \beta}  u^{\alpha,\beta} \sum_{R} S_{\alpha,R}^z S_{\beta,R}^z \Big).
\end{equation}

Here, the correlator involves the $z$ component of the spin $S_{\alpha,R}^z$ only for different orbitals on site $R$, as suggested by the definition of the Hund's coupling.

The uncorrelated state $|\Phi_0\rangle$ is now the ground state of a new quadratic auxiliary Hamiltonian, which is appropriate for the multi-orbital model. Focusing only on the investigation of superconductivity in the model and neglecting any possible magnetic order, the auxiliary Hamiltonian can be defined as:
\begin{equation}
    \mathcal{H}_{aux}=\mathcal{H}_{0}+\mathcal{H}_{BCS},
\end{equation}

where $\mathcal{H}_0$ is the same as the tight-binding Hamiltonian of Eq.~(\ref{eq:TB}), while $\mathcal{H}_{BCS}$ is now defined as:

\begin{equation}
 \mathcal{H}_{BCS} = - \sum_{R,\alpha,\sigma} \mu_{\alpha} c_{R,\alpha,\sigma}^\dag c_{R,\alpha,\sigma} 
    + \sum_{R,\alpha,\delta} \left[ \Delta_{\alpha,\delta} \left( c_{R,\alpha,\uparrow} c_{R+\delta,\alpha,\downarrow}
     - c_{R,\alpha,\downarrow} c_{R+\delta,\alpha,\uparrow} \right) + \textrm{h.c.}  \right],
\label{eq:H_aux}
\end{equation}
where $\delta=x$, $y$, $x+y$, and $x-y$ indicates nearest and next-nearest neighbors of the site $R$ and $\mu_{\alpha}$ defines the chemical potential 
of orbital $\alpha$. The (singlet) pairing amplitudes $\Delta_{\alpha,\delta}$ and $\mu_{\alpha}$ are optimized, while the hopping parameters in $\mathcal{H}_{0}$ are usually kept fixed to the bare Hamiltonian ones. Inter-orbital pairing amplitudes are found to be negligible in the optimal wave function, in 
agreement with~\cite{misawa2014} and therefore not included in the auxiliary Hamiltonian.

\section{Optimizing the variational state: Stochastic Reconfiguration}\label{sec:SR}

\index{stochastic reconfiguration} We have introduced the functional form of the wave function, which depends on many variational parameters, ranging from the pseudopotentials in the Jastrow factor to the BCS coupling or the hopping parameters in the auxiliary Hamiltonian. This freedom in the choice of the variational wave function is meaningful only if we can construct a procedure that allows the optimization of many variational parameters at the same time, in order to find the best approximation of the ground state. One possible way to optimize the set of parameters is to use the zero-variance property, minimizing the variance of the energy in the variational state~\cite{umrigar1988}. In contrast, the \textit{stochastic reconfiguration} exploits the minimization of the variational energy; unlike the minimization of variance, where the minimum value is zero, the ground state energy is unknown a priori~\cite{Sorella2005}. However, for a given form of the trial wave function, energy-minimized wave functions yield more accurate values of expectation values than wave functions whose variance is minimized~\cite{snajdr2000}.

Let us explain the stochastic reconfiguration assuming a generic functional form for the wave function $\ket{\Psi(\{p_k\})}$ with $k = 1 \ldots N_p$, which depends on a set of $N_p$ variational parameters $\{p_k\}$. The variational energy of the state can be decreased by the application of the projection operator $(\mathcal{H}-\Lambda)$, where $\Lambda$ is an appropriate constant energy shift:

\begin{equation}
    \ket{\Psi_\Lambda} = (\mathcal{H} - \Lambda) \ket{\Psi^0},
    \label{3.41}
\end{equation}

where $\ket{\Psi^0} = \ket{\Psi(\{p_k^0\})}$ is the starting state, depending on an initial set of variational parameters $\{p_k^0\}$. However, in general, the new state $ \ket{\Psi_\Lambda}$ will not be written in the Jastrow-Slater form. Therefore, the idea is to find the quantum state $\ket{\Psi'} = \ket{\Psi(\{p_k'\})}$, which gives the best possible approximation of $ \ket{\Psi_\Lambda}$. If the change of the variational parameters is taken to be small, we can approximate the new parameters as $p_k' = p^0_k + \delta p_k$, and the new state $\ket{\Psi'}$ can be expanded at first order in $\{\delta p_k\}$:

\begin{equation}
    \ket{\Psi'} = \delta p_0 \ket{\Psi^0} + \sum_{k=1}^{N_p} \delta p_k \frac{\partial \ket{\Psi^0}}{\partial p_k} + O(\delta p_k^2).
\end{equation}

Here, the series is truncated to the second order, and $\delta p_0$ is considered to match the normalization condition. Then, we define the local operator $\mathcal{O}_k$ (with $k=1,\dots,N_p$) on the configuration $\ket{x}$ as the logarithmic derivative with respect to the variational parameter $p_k$:

\begin{equation}
    \mathcal{O}_k = \frac{1}{\Psi^0(x)} \frac{\partial \Psi^0(x)}{\partial p_k} = \frac{\partial \ln{\langle x|\Psi^0\rangle} }{\partial p_k}.
\end{equation}

Imposing $\mathcal{O}_0 = 1$, $\ket{\Psi'}$ can be expressed by using the logarithmic derivative operators:

\begin{equation}
    \ket{\Psi'} = \sum_{k=0}^{N_p} \delta p_k \mathcal{O}_k \ket{\Psi^0}.
    \label{3.44}
\end{equation}

Then, the stochastic reconfiguration method imposes that the projection of the shifted state $\ket{\Psi_\Lambda}$ and of the new state $\ket{\Psi'}$ on the subspace $\mathcal{O}_j\ket{\Psi^0}$ is equal for $j = 0 \ldots N_p$:

\begin{equation}
    \bra{\Psi^0} \mathcal{O}_j \ket{\Psi'} = \bra{\Psi^0} \mathcal{O}_j \ket{\Psi_{\Lambda}}.
\end{equation}

This ensures that the energy computed on $|\Psi'\rangle$ is lower than the one computed on $|\Psi^0\rangle$.
Then, by inserting the definition of $\ket{\Psi_\Lambda}$ from Eq.~(\ref{3.41}) and the one of  $\ket{\Psi'}$ from Eq.~(\ref{3.44}), the stochastic reconfiguration condition becomes:

\begin{equation}
    \sum_{k=0}^{N_p} \delta p_k \langle \mathcal{O}_j \mathcal{O}_k \rangle = \langle \mathcal{O}_j (\Lambda - \mathcal{H}) \rangle.
\end{equation}

The symbol $\langle \ldots \rangle$ stands for $\bra{\Psi^0} \ldots \ket{\Psi^0}$. Subsequently, the conditions for $j=0$ and $j \neq 0$ are separated, and the $k=0$ term is extracted from the summation, leading to the following relations:

\begin{gather}
    \delta p_0 = - \sum_{k=1}^{N_p} \delta p_k \langle \mathcal{O}_k \rangle + \Lambda - \langle \mathcal{H} \rangle \nonumber \\ 
    \delta p_0 \langle \mathcal{O}_j \rangle + \sum_{k=1}^{N_p} \delta p_k \langle \mathcal{O}_j \mathcal{O}_k \rangle = \Lambda \langle \mathcal{O}_j \rangle - \langle \mathcal{O}_j \mathcal{H} \rangle.
\end{gather}
    
The parameter $\delta p_0$, related to normalization, does not affect any physical observable; indeed, the two conditions can be unified in a compact form, in the unknowns $\{\delta p_k\}$ as:

\begin{equation}
    \sum_{k = 1}^{N_p} \delta p_k S_{jk} = f_j,
\end{equation}

where $f_j = \langle \mathcal{O}_j \rangle \langle \mathcal{H} \rangle - \langle \mathcal{O}_j \mathcal{H} \rangle$ is the generalized force and $S_{jk} = \langle \mathcal{O}_j \mathcal{O}_k \rangle - \langle \mathcal{O}_j \rangle \langle \mathcal{O}_k \rangle$ is the covariance matrix of logarithmic derivatives of $N_p \times N_p$ dimension (symmetric and semi-positive definite). During the Monte Carlo procedure, the expectation values are computed by stochastic sampling, the linear system is numerically solved, and, successively, parameters are updated according to the following scheme:

\begin{gather}
    \delta p_k  =\sum_{j = 1}^{N_p} S_{jk}^{-1} f_j \nonumber \\
    p_k' = p_k^0 + \tau \delta p_k.
    \label{3.49}
\end{gather}

$\tau$ is the acceleration parameter that determines the convergence of the algorithm (in principle, each parameter could have a different $\tau_k$). A large $\tau$ guarantees rapid convergence, although it could threaten the stability of the algorithm. Hence, $\tau$ is chosen sufficiently small to generate a variational energy that can be expanded in series:

\begin{equation}
    E(\Psi') = E(\Psi_0) + \tau \sum_{k=0}^{N_p} \frac{\partial E(\Psi_0)}{\partial p_k} \delta p_k + O(\tau^2).
\end{equation}

Therefore, by using the definition of $\delta p_k$ in Eq.~(\ref{3.49}) and the fact that the derivative of the energy with respect to a parameter is equal to minus the generalized force ($\frac{\partial E(\Psi_0)}{\partial p_k} = - f_k$), we can rewrite the change in the variational energy as:

\begin{equation}
     \delta E = E(\Psi') - E(\Psi_0) = - \tau \sum_{j,k = 1}^{N_p} S_{jk}^{-1} f_j f_k + O(\tau^2).
\end{equation}

Here, the decrease of the variational energy ($\delta E \leq 0$) is ensured by the semi-positive definition of the covariance matrix of the logarithmic derivatives $S_{jk}$. 

Examples of calculations of logarithmic derivative operators for Jastrow-Slater wave functions can be found in Refs.~\cite{Becca2017,yunoki2006}

\section{Correlation functions in Variational Monte Carlo: Definitions and examples}\label{sec:results}

In this section, we introduce a few relevant correlators, which expectation values over the optimal variational state allow us to describe the properties of our variational approximation of the ground-state wave function. 

\index{static structure factor} The first one is the \emph{static structure factor} $N({\bf q})$ defined in reciprocal space as:
\begin{equation}\label{eq:nqnq}
N({\bf q}) = \langle n_{\mathbf{q}}n_{-\mathbf{q}}\rangle=\frac{1}{L} \sum_{R,R^\prime}\langle n_{R} n_{R^\prime} \rangle e^{i{\bf q}\cdot({\bf R}-{\bf R^\prime})},
\end{equation}
where $\langle \dots \rangle$ indicates the expectation value over the variational wave function. In particular, the presence of a peak at a given ${\bf q}$ vector, diverging with the system size, denotes the presence of true charge order in the system. A nondiverging peak signals the presence of short-range charge correlations that do not lead to long-range order.

Moreover, the static structure factor allows us to assess the metallic or insulating nature of the ground state by considering just ground-state expectation values, without directly calculating
energy differences. Indeed, the behavior of the static structure factor at small momenta (or, equivalently, at large wavelengths) reveals the nature of the state associated with that wave function. To show this, let us recall the Feynman single-mode approximation, originally formulated in the context of collective excitations in liquid Helium~\cite{feynman1954} and then applied to fermionic systems~\cite{overhauser1971}. In this context, a variational Ansatz of an excited state $|\Psi_{\mathbf{q}}\rangle$, with a given momentum $\mathbf{q}$, can be obtained by applying $n_{\mathbf{q}}$ to the ground state wave function $|\Psi\rangle$ 
(or one approximation for it), namely:
\begin{equation}\label{eq:excit}
|\Psi_{\mathbf{q}}\rangle = n_{\mathbf{q}}|\Psi\rangle.
\end{equation}
The variational estimator of the excitation energy is then given by:

\begin{equation}
E_{\mathbf{q}}-E_0=\frac{\langle \Psi_{\mathbf{q}}|({\cal H}-E_0)|\Psi_{\mathbf{q}}\rangle}{\langle \Psi_{\mathbf{q}}|\Psi_{\mathbf{q}}\rangle}=
\frac{\langle \Psi|n_{-\mathbf{q}}[{\cal H},n_{\mathbf{q}}]|\Psi\rangle}{\langle \Psi_{\mathbf{q}}|\Psi_{\mathbf{q}}\rangle}=\frac{\langle \Psi|[n_{-\mathbf{q}}{\cal H}],n_{\mathbf{q}}|\Psi\rangle}{\langle \Psi_{\mathbf{q}}|\Psi_{\mathbf{q}}\rangle},
\end{equation}
where $\mathcal{H}$ is the Hubbard Hamiltonian. The sum of both commutators 
$(n_{-\mathbf{q}}{\cal H}n_{\mathbf{q}}-n_{-\mathbf{q}}n_{\mathbf{q}}{\cal H})+(n_{-\mathbf{q}}{\cal H}n_{\mathbf{q}}-{\cal H}n_{-\mathbf{q}}n_{\mathbf{q}})$
is equivalent to the double commutator 
\begin{equation}
[n_{-\mathbf{q}},[{\cal H},n_{\mathbf{q}}]]=
n_{-\mathbf{q}}({\cal H} n_{\mathbf{q}}-n_{\mathbf{q}}{\cal H})-({\cal H} n_{\mathbf{q}}-n_{\mathbf{q}}{\cal H})n_{-\mathbf{q}}
\end{equation}
due to the inversion symmetry $\mathbf{q} \leftrightarrow -\mathbf{q}$. 
Therefore, the excitation energy is given by
\begin{equation}
E_{\mathbf{q}}-E_0=\frac{1}{2}\frac{\langle\Psi| [n_{-\mathbf{q}},[{\cal H},n_{\mathbf{q}}]]|\Psi\rangle}{N_{\mathbf{q}}},
\end{equation}
where $N(\mathbf{q})=\langle \Psi|n_{-\mathbf{q}}n_{\mathbf{q}}|\Psi \rangle$ is the static structure 
factor for the ground state. The double commutator $[n_{-\mathbf{q}},[{\cal H},n_{\mathbf{q}}]]$ 
can be evaluated and involves the kinetic term only, since the potential term of the Hamiltonian contains density-density interactions that commute with $n_{\mathbf{q}}$~\cite{tocchio2011}:
\begin{equation}
[n_{-\mathbf{q}},[{\cal H},n_{\mathbf{q}}]]= \frac{1}{L} \sum_{\mathbf{k},\sigma} 
(\epsilon_{\mathbf{k}+\mathbf{q}}+\epsilon_{\mathbf{k}-\mathbf{q}} - 2 \epsilon_{\mathbf{k}}) 
c^\dag_{\mathbf{k},\sigma} c^{\phantom{\dagger}}_{\mathbf{k},\sigma}, 
\end{equation}
where $\epsilon_{\mathbf{k}}$ is the dispersion relation of the kinetic part of the Hubbard Hamiltonian. Then, in the limit of small momenta, one gets:
\begin{equation}\label{eq:gap}
E_{\mathbf{q}}-E_0 \propto \lim_{q\to 0} \frac{|\mathbf{q}|^2}{N(\mathbf{q})}.
\end{equation}
Then, whenever $N(\mathbf{q})$ is linear in $\mathbf{q}$, the energy spectrum is gapless, suggesting that the system is metallic; by contrast, if $N(\mathbf{q})\sim \mathbf{q}^2$, this construction gives a finite gap, leading to an insulator.

\begin{figure}[t!]
 \centering
 \includegraphics[width=0.8\textwidth]{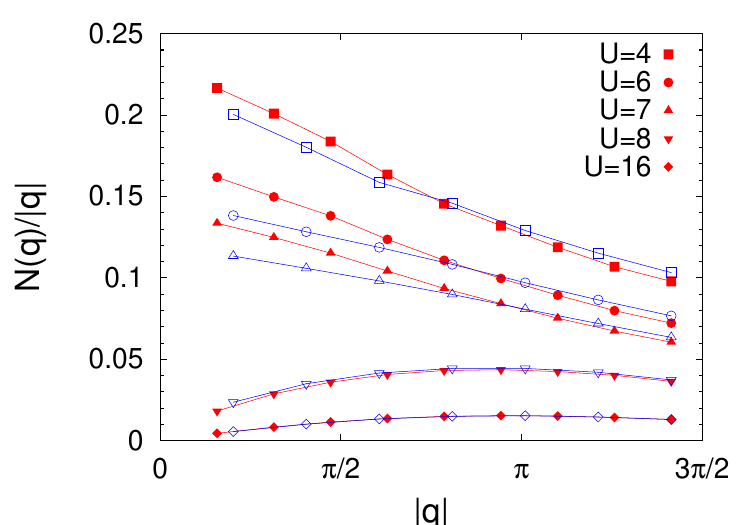}
 \caption{Variational results for $N(\mathbf{q})$ divided by $|\mathbf{q}|$, as a function of $|\mathbf{q}|$, for $L=98$ (empty symbols) and for $L=162$ (full
symbols) at $t'/t = 0.75$. The metal-insulator transition takes place between $U/t = 7$ and $U/t = 8$, that is, where $N(\mathbf{q})$ changes its behavior from linear to quadratic for $|\mathbf{q}|\to 0$~\cite{tocchio2008}.}
 \label{fig:Nq}
\end{figure}

As an example, we show in Fig.~\ref{fig:Nq}, the behavior of $N(\mathbf{q})$ across the metal-insulator transition that occurs in the single-orbital Hubbard model as a function of $U/t$ at $t'/t=0.75$. In order to highlight the location of the metal-insulator transition we plot $N(\mathbf{q})/|\mathbf{q}|$: If the system is metallic, $N(\mathbf{q})/|\mathbf{q}|$ extrapolates to a finite value for $|\mathbf{q}|\to 0$; conversely, if the system is insulating, $N(\mathbf{q})/|\mathbf{q}|$ extrapolates to zero for $|\mathbf{q}|\to 0$; see Eq.~(\ref{eq:gap}).

\index{spin-spin correlations} The second one is the spin-spin correlator. Analogously to the case of charge order, spin order in the system is associated to a peak in the \emph{spin-spin correlations} defined as:
\begin{equation}\label{eq:sqsq}
S({\bf q}) = \frac{1}{L} \sum_{R,R^\prime}\langle S^z_{R} S^z_{R^\prime} \rangle e^{i{\bf q}\cdot({\bf R}-{\bf R^\prime})},
\end{equation}
where $S^z_{R}$ is the spin operator along the $z$ direction, i.e., $S^z_{R}=1/2(c^{\dagger}_{R,\uparrow} c^{\phantom{\dagger}}_{R,\uparrow} -
 c^{\dagger}_{R,\downarrow} c^{\phantom{\dagger}}_{R,\downarrow})$. Also, the presence of a spin gap can be detected by looking at the behavior of $S(\mathbf{q})$ at small momenta.

\begin{figure}[t!]
 \centering
 \includegraphics[width=0.8\textwidth]{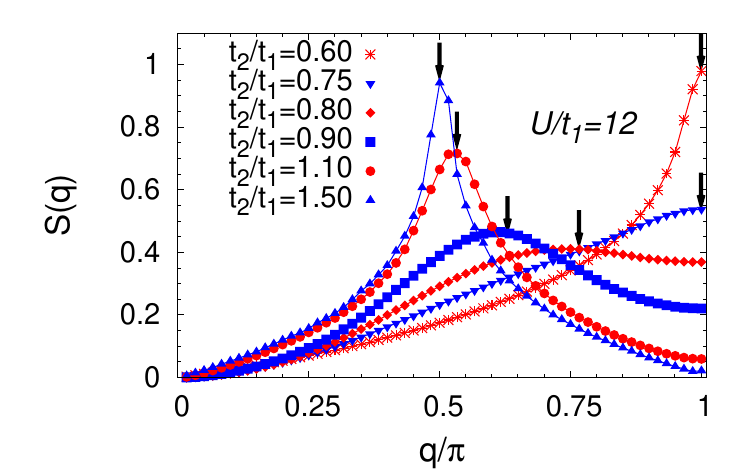}
 \caption{Spin-spin correlations $S(\mathbf{q})$ at $U/t_1 = 12$ for different values of the hopping ratio $t_2/t_1$. Arrows indicate the location of the peaks. Data are shown on a $L=120$ lattice size~\cite{tocchio2010}.}
 \label{fig:Sq}
\end{figure}

 As an example, we show the behavior of $S(\mathbf{q})$ for the one dimensional Hubbard model. The model is the same as the one we introduced in Eq.~(\ref{eq:Hubbard_single}), just denoting nearest neighbor hopping as $t_1$ and next-nearest neighbor hopping as $t_2$. Even if in one dimension there is no long-range order, short-range spin correlations are present in the insulating phase. They can be detected as a nondiverging peak in $S(\mathbf{q})$. The peak remains commensurate at $Q=\pi$ up to $t_2/t_1\sim 0.75$ and then shifts,  as a function of the ratio $t_2/t_1$, to the value of $Q=\pi/2$ for large values of $t_2/t_1$. Results can be seen in Fig.~\ref{fig:Sq}.

 \index{pairing correlation function} Then, another crucial task in ground-state simulations is the detection of superconductivity in the model. The presence of finite BCS parameters in the variational state after optimization is a necessary condition for the presence of superconductivity. It is not a sufficient condition because either the system can be an insulator, induced by the long-range tail of the Jastrow factor in the variational state, or finite BCS parameters can be a finite-size effect, disappearing in the thermodynamic limit, as it occurs in the $U=0$ limit of the Hubbard model. Hence, to reveal the presence of true superconductivity in the state, we can compute the so-called \textit{singlet pairing correlation function} in real space, that we restrict here to be intra-orbital:
\begin{equation}
    D_{\alpha}(r) = \frac{1}{L} \sum_R \langle P_{\alpha,R} P_{\alpha,R+r}^\dag \rangle,
\end{equation}

where $\langle \ldots \rangle$ means the usual expectation value on the variational ground state, $\alpha$ is the orbital index, and the singlet operator is defined as: $P_{\alpha,R} = c_{\alpha,R+\eta,\downarrow} c_{\alpha,R,\uparrow} - c_{\alpha,R+\eta,\uparrow} c_{\alpha,R,\downarrow}$. In practice, the $P_{\alpha,R}$ operator destroys a singlet  oriented along the $\eta$ direction (for instance, $\eta=x$ or $\eta=y$) in position $R$, while $P_{\alpha,R+r}^\dag$ creates a singlet with the same orientation in position $R + r$. Then, the system is superconductive when the following condition is met:
\begin{equation}
     \phi_\alpha^2 = \lim_{r \to \infty} D_{\alpha}(r) \; \text{is finite}.
\end{equation}

The symbol $\phi_\alpha$ denotes the superconductive order parameter, that, for $|\phi_\alpha| \neq 0$ indicates that the system is superconductive.

From a practical point of view, on a finite lattice, we evaluate $D_{\alpha}(r)$ at the maximum distance, i.e. $r = l/2$ (if the lattice size is defined as $L = l \times l$) because of the periodic boundary conditions. Then, we consider different lattice sizes and make an extrapolation of $\phi_{\alpha}$ to the thermodynamic limit ($l \to \infty$), in order to assess the actual presence of superconductivity. Moreover, it might be useful to compare the superconductive correlation functions for finite values of $U$ with the ones obtained at $U = 0$. In the latter case, finite-size kinetic contributions from the hopping are present. Hence,  superconductivity emerges only when we observe an increase of $\phi_\alpha$ in the presence of interaction, with respect to the $U=0$ case.

\begin{figure}[t!]
 \centering
 \includegraphics[width=1.0\textwidth]{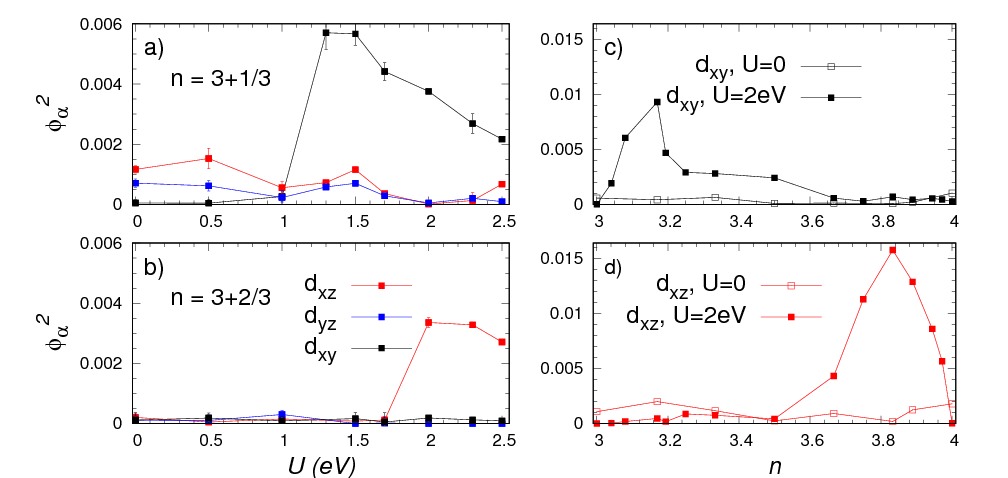}
 \caption{Square of the superconducting order parameter $\phi_{\alpha}^2$ in each orbital as a function of $U$ at $n=3+1/3$ [panel a)] and $n=3+2/3$ [panel b)]. Data are for the $L=18\times 18$ cluster. Square of the superconducting order parameter in the $d_{xy}$ [panel c)] and $d_{xz}$ [panel d)] orbitals as a function of $n$ at $U=2$eV (full squares) and at $U=0$ (empty squares). Data are for the $L=12\times 12$ cluster. All data are computed at $J=0.2U$~\cite{marino2025}.}
 \label{fig:Superc}
\end{figure}

As an example, we show in Fig.~\ref{fig:Superc}, left panels, the results for $\phi_{\alpha}^2$ as a function of $U$ in the three-orbital Hubbard-Kanamori Hamiltonian of Eqs.~(\ref{eq:HubbardKanamori}) and (\ref{eq:TB}). Results are presented for two relevant electronic densities of $n=3+1/3$ and $n=3+2/3$. In both cases, superconductivity develops only in one of the three orbitals, while the other two orbitals, instead, do not show any enhancement of the superconducting correlations with respect to the $U=0$ case. We also show the behavior 
of $\phi_{\alpha}^2$ as a function of density $n$ in Fig.~\ref{fig:Superc}, right panels, for a value of $U$ where superconductivity is present in one of the orbitals. Here, we also report the non-interacting values at $U=0$, to emphasize the relevance of the Hubbard interaction to generate the electron pairing. Even if not shown here, we mention that superconductivity is not present when the Hund's coupling is set to zero.

\begin{figure}[t!]
 \centering
 \includegraphics[width=1.0\textwidth]{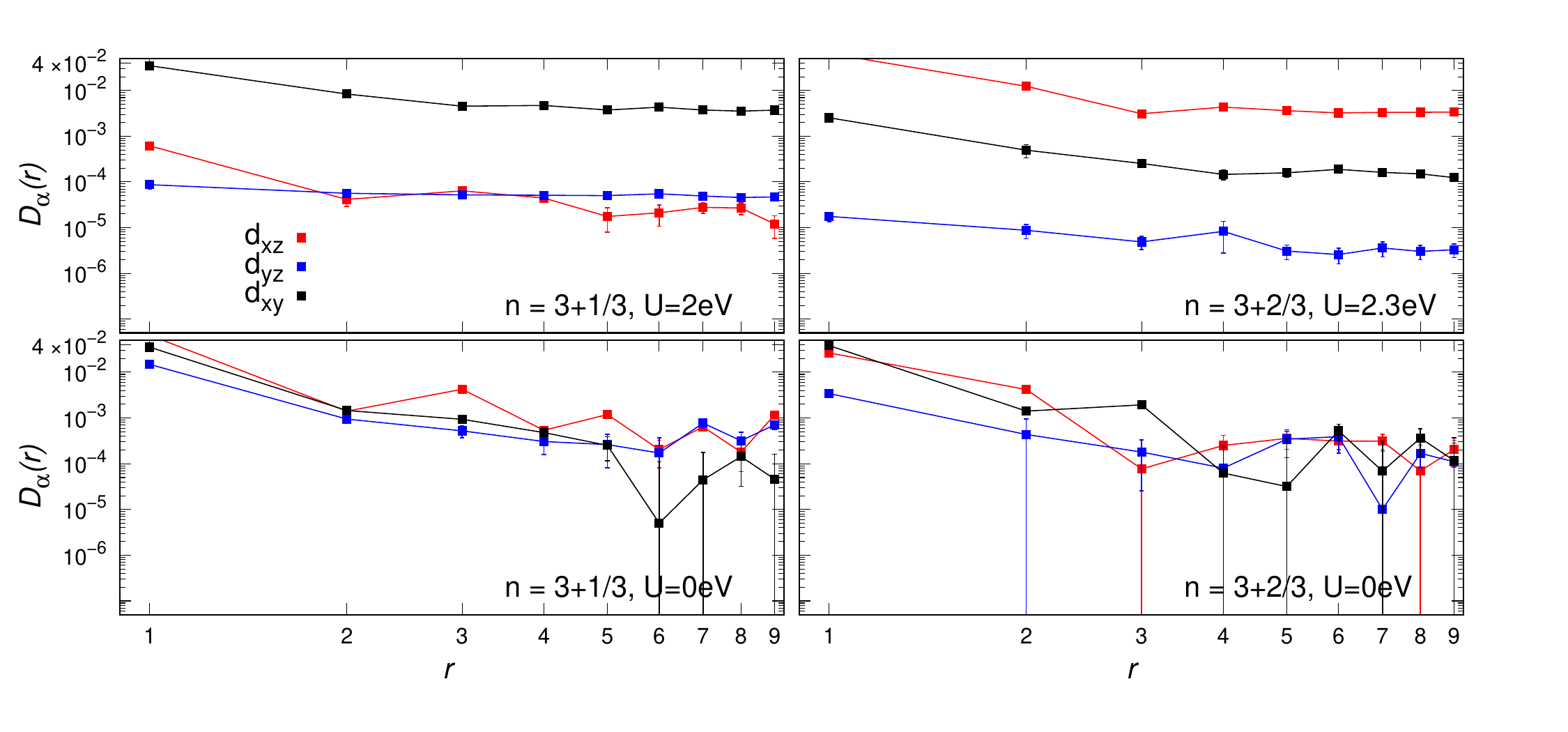}
 \caption{Upper panels: $D_{\alpha}(r)$ in all orbitals, for a value of $U$ where superconductivity is present. Data are shown at electronic densities $n=3+1/3$ (left column) and $n=3+2/3$ (right column). Lower panels: The same as above, but for $U=0$. Data are computed on a $L=18\times18$ cluster at $J=0.2U$~\cite{marino2025}.}
 \label{fig:Superc_U0}
\end{figure}

As a comparison, in Fig.~\ref{fig:Superc_U0}, we report the intra-orbital pairing correlations $D_{\alpha}(r)$ at densities $n=3+1/3$ and $n=3+2/3$, both within the orbital-selective phase and at $U=0$. In both cases, there is one orbital where pairing correlations at large distances are strongly enhanced going from $U=0$ to large values of $U$. On the contrary, in the other two orbitals, pairing correlations remain similar or are even suppressed when electron-electron correlation is switched on. 

\begin{figure}[t!]
 \centering
 \includegraphics[width=1.0\textwidth]{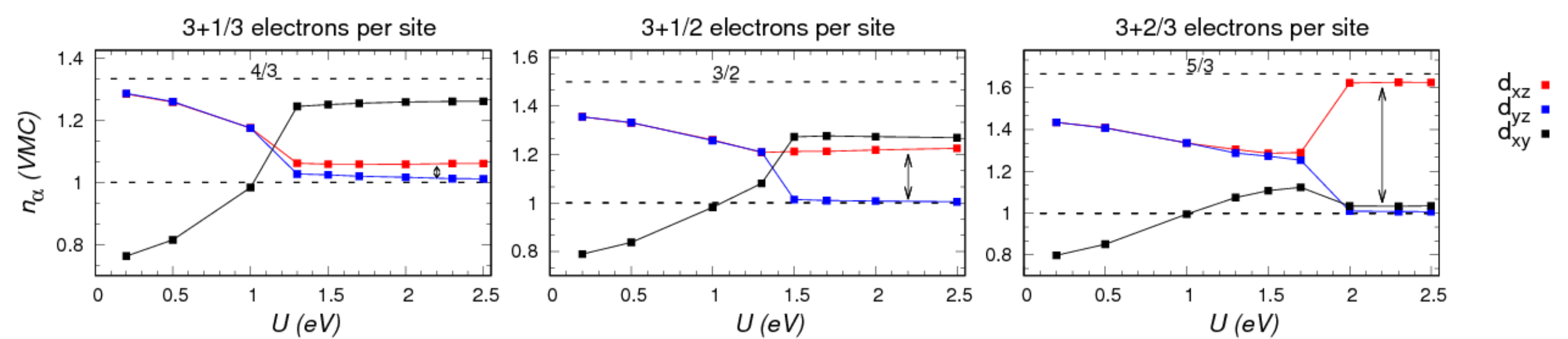}
 \caption{Electronic density per orbital $n_\alpha$ as a function of $U$ for three
different total electronic densities. We
fix $J/U = 0.2$ and compute data for the $L = 12 \times 12$ cluster. The arrows highlight the size of the differentiation between the $d_{xz}$ and the $d_{yz}$ orbitals~\cite{marino2025}.}
 \label{fig:orbital}
\end{figure}

 \index{orbital selectivity} We mention, that in addition to the correlators introduced above, one can also compute expectation values of simpler single-particle observables. For instance, in a multi-orbital model, one could compute the electronic density per orbital, defined as $n_\alpha=\frac{1}{L}\langle \sum_{R,\sigma} n_{R,\alpha,\sigma}\rangle$. This observable allows us to identify orbital selective phenomena, that are related to superconductivity. In Fig.~\ref{fig:orbital}, we observe a differentiation in the orbital occupation of the two degenerate orbitals ($d_{xz}$ and $d_{yz}$) above a critical value of $U$ that corresponds to the onset of superconductivity in the most occupied orbital. We remark that the onset of superconductivity and orbital selectivity is related to the presence of a long-range tail in the Jastrow term of the variational wave function, that allows us for a proper treatment of spatial correlations. 

\section{Final remarks}\label{sec:final}

In this lecture, we introduced the basics of the variational Monte Carlo method in general and in relation with the Hubbard model, both in its simpler single-orbital version and in its multi-orbital extension. 

Variational Monte Carlo is a conceptually simple approach, where the computation of physical observables over a variational approximation of the ground state wave function is rewritten as averages of local observables over the usual quantum mechanical probability distribution. This definition ensures that VMC has no sign problem, since we sample a nonnegative quantity.

The core of the variational Monte Carlo approach is to define accurate variational states that can reproduce the correct ground state physics; for this reason, we devoted Sec.~\ref{sec:wf} to the definition of different variational states. We described the Jastrow-Slater wave function for the single-orbital Hubbard model and how it can be improved with the backflow terms. We also presented an extension of the Jastrow-Slater wave function to the multi-orbital case.

Then, once a functional form of the variational wave function is built, it is crucial to be able to optimize the variational parameters that are contained within the variational state. The concomitant optimization of many variational parameters is made possible with the stochastic reconfiguration method, described in Sec.~\ref{sec:SR}. 

Finally, we introduced several may-body correlators that allow us to characterize the physics of the optimal variational approximation of the ground state. Among the other, we discussed how to discriminate whether the system is metallic or insulating, detect the presence of charge or spin order and the emergence of superconductivity. All these features lead to a complete understanding of the ground-state properties of our correlated Hubbard Hamiltonian.  

We close this section by commenting on which paths might variational Monte Carlo take in the future. 

A first one is based on the application of relatively simple variational states, of the Jastrow-Slater form, to complicated multi-orbital Hamiltonians where approaches that are nonperturbative in electronic correlations and can address spatial correlations are lacking. These ideas have been already discussed in these lecture notes, in the field of iron-based superconductors, and can have a relevant impact, for instance, in the field of Kagome metals. Here, electronic correlations are a necessary ingredient, while the complicated kinetic electronic model is a challenge for all the contemporary many-body methods~\cite{disante2026}.

A second one is the improvement of variational wave functions with respect to the scheme introduced in this lecture. In this respect, neural-network quantum states, based on the transformer architecture, show excellent accuracy and appear as promising tools for solving the intricacies of competing orders in the Hubbard model. However, the domain of applicability of these more evolved variational wave functions is, at the moment, limited to the single-orbital Hubbard model on latices with a simple geometry~\cite{moreno2022,rende2026,viteritti2026}. 

Finally, a third one that goes away from the target of these notes is to consider not only ground-state properties, but to provide VMC calculations of dynamics and finite-temperature properties, that involve also excited states. This is a wide growing field, based on the time-dependent variational principle, which might be the subject of another set of lecture notes.


\clearchapter

\clearpage
\begin{thebibliography}{99}
 \bibitem{Becca2017} F.~Becca and S.~Sorella: {\em Quantum Monte Carlo Approaches for Correlated Systems} \\ (Cambridge University Press, 2017).
 \bibitem{Schollwoeck2011} U.~Schollw\"ock, The density-matrix renormalization group in the age of matrix product states, Ann.~Phys.~\textbf{326}, 96 (2011)
 \bibitem{georges1996} A.~Georges, G.~Kotliar, W.~Krauth, and M.~J.~Rozenberg, Dynamical mean-field theory of strongly correlated fermion systems and the limit of infinite dimensions, Rev.~Mod.~Phys.~\textbf{68}, 13 (1996)
 \bibitem{Sorella2005} S.~Sorella, Wave function optimization in the variational Monte Carlo method, Phys.~Rev.~B~\textbf{71}, 241103 (2005)
 \bibitem{Leblanc2015} J.~P.~F.~LeBlanc {\em et al.}, Solutions of the Two-Dimensional Hubbard Model: Benchmarks and Results from a Wide Range of Numerical Algorithms, Phys.~Rev.~X~\textbf{5,} 041041 (2015)
\bibitem{qin2022} M.~Qin, T.~Sch\"afer, S.~Andergassen, P.~Corboz, and E.~Gull, The Hubbard model: A computational perspective, Annu.~Rev.~Conden.~Matter Phys.~\textbf{13}, 275 (2022)
\bibitem{pavarini2001} E.~Pavarini, I.~Dasgupta, T.~Saha-Dasgupta, O.~Jepsen, and O.~Andersen, Bandstructure trend in hole-doped cuprates and correlation with $T_{c\; \textrm{max}}$, Phys.~
Rev.~Lett.~\textbf{87}, 047003 (2001)
\bibitem{lieb1968} E.~H.~Lieb, and F.~Y.~Wu, Absence of Mott transition in an exact solution of the short-range, one-band model in one dimension, Phys.~Rev.~Lett.~\textbf{20}, 1445 (1968)
\bibitem{zhang1995} S.~Zhang, J.~Carlson, and J.~E.~Gubernatis, Constrained Path Quantum Monte Carlo Method for Fermion Ground States, Phys.~Rev.~Lett.~\textbf{74}, 3652 (1995)
\bibitem{kanamori1963} J.~Kanamori, Electron correlation and ferromagnetism of transition metals, Progr.~Theor.~Phys.~\textbf{30}, 275 (1963)
\bibitem{georges2013} L.~Georges, L.~ de Medici and J.~Mravlje,  Strong correlations from Hund’s
coupling,  Annu.~Rev.~Condens.~Matter Phys.~\textbf{4}, 137 (2013)
\bibitem{daghofer2010} M.~Daghofer, A.~Nicholson, A.~Moreo, and E.~Dagotto, Three orbital model
for the iron-based superconductors, Phys.~Rev.~B \textbf{81}, 014511 (2010)
\bibitem{fanfarillo2020} L.~Fanfarillo, A.~Valli, and M.~Capone, Synergy between Hund-driven correlations and boson-mediated superconductivity, Phys.~Rev.~Lett.~\textbf{125}, 177001 (2020)
 \bibitem{zheng2017} B.-X.~Zheng {\em et al.}, Stripe order in the underdoped region of the two-dimensional Hubbard model, Science~\textbf{358}, 1155 (2017)
\bibitem {darmawan2018} A.~S.~Darmawan, Y.~Nomura, Y.~Yamaji, and M.~Imada, Stripe and superconducting order competing in the Hubbard model on a square lattice studied by a combined variational Monte Carlo and tensor network method, Phys.~Rev.~B \textbf{98}, 205132 (2018)
\bibitem{marino2022} V.~Marino, F.~Becca, and L.~F.~Tocchio, Stripes in the extended $t-t'$ Hubbard model: A Variational Monte Carlo analysis, SciPost Phys.~\textbf{12}, 180 (2022) 
\bibitem{capello2005} M.~Capello, F.~Becca, M.~Fabrizio, S.~Sorella, and E.~Tosatti, Variational Description of Mott Insulators, Phys.~Rev.~Lett.~\textbf{94}, 026406 (2005)
\bibitem{anderson1987} P.~W.~Anderson, The resonating valence bond state in La$_2$CuO$_4$ and superconductivity,  Science \textbf{235}, 1196 (1987)
\bibitem{gutzwiller1963} M.~C.~Gutzwiller, Effect of correlation on the ferromagnetism of transition
metals. Phys.~Rev.~Lett.~\textbf{10}, 159 (1963)
\bibitem{tocchio2008} L.~F.~Tocchio, F.~Becca, A.~Parola, and S.~Sorella, Role of backflow correlations for the non-magnetic phase of the $t-t'$ Hubbard model, 
Phys.~Rev. B \textbf{78}, 041101(R) (2008)
\bibitem{tocchio2011} L.~F.~Tocchio, F.~Becca, and C.~ Gros, Backflow correlations in the Hubbard model: an efficient tool for the metal-insulator transition and the large-$U$ limit, Phys.~Rev.~B \textbf{83}, 195138 (2011)
\bibitem{tocchio2013} L.~F.~Tocchio, H.~Feldner, F.~ Becca, R.~ Valenti, and C.~ Gros, Spin-liquid versus spiral-order phases in the anisotropic triangular lattice, Phys.~Rev.~B \textbf{87}, 035143 (2013)
\bibitem{chitra1995} R.~Chitra, S.~Pati, H.~R.~Krishnamurty, D.~Sen, and S.~Ramasesha, Density-matrix renormalization-group studies of the spin-1/2 Heisenberg system with dimerization and frustration, Phys.~Rev.~B \textbf{52}, 6581 (1995)
\bibitem{luo2019} D.~Luo and B.~K.~Clark, Backflow Transformations via Neural Networks for Quantum Many-Body Wave Functions, Phys.~Rev.~Lett.~\textbf{122}, 226401 (2019)
\bibitem{marino2025} V.~Marino, A.~Scazzola, F.~Becca, M.~Capone, and L.~F.~Tocchio, Intertwined Superconductivity and Orbital Selectivity in a Three-Orbital Hubbard Model for the Iron Pnictides, Phys.~Rev.~Lett.~\textbf{134}, 196502 (2025)
\bibitem{marino2025b} V.~Marino, Superconductivity and Orbital Selectivity in a three-orbital Hubbard model for the iron-based superconductors - Variational Monte Carlo analysis, Doctoral Thesis, URL https://iris.polito.it/handle/11583/3001278 (2025)
\bibitem{misawa2014} T.~Misawa and M.~Imada, Superconductivity and its mechanism in an \emph{ab initio} model for electron-doped LaFeAsO, Nat.~Comm.~\textbf{5}, 5738 (2014)
\bibitem{umrigar1988} C.~J.~Umrigar, K.~G.~Wilson, and J.~W.~Wilkins. Optimized trial wave functions
for quantum Monte Carlo calculations, Phys.~Rev.~Lett.~\textbf{60}, 1719 (1988)
\bibitem{snajdr2000} M.~Snajdr and S.~M.~Rothstein, Are properties derived from variance-optimized wave functions generally more accurate? Monte Carlo study of non-energy-related properties of H$_2$, He, and LiH, J.~Chem.~Phys.~\textbf{112}, 4935(2000)
\bibitem{yunoki2006} S.~Yunoki and S.~Sorella, Two spin liquid phases in the spatially anisotropic triangular Heisenberg model, Phys.~Rev.~B \textbf{74}, 014408 (2006)
\bibitem{feynman1954} R.~P.~Feynman, Atomic Theory of the Two-Fluid Model of Liquid Helium, Phys.~Rev.~{\bf 94}, 262 (1954)
\bibitem{overhauser1971} A.~W.~Overhauser, Simplified Theory of Electron Correlations in Metals, Phys.~Rev.~B \textbf{3}, 1888 (1971)
\bibitem{tocchio2010} L.~F.~Tocchio, F.~Becca, and C.~Gros, Interaction induced Fermi-surface renormalization in the $t_1-t_2$ Hubbard model close to the Mott-Hubbard transition, Phys.~Rev.~B \textbf{81}, 205109 (2010)
\bibitem{disante2026} D.~Di Sante, T.~Neupert, G.~Sangiovanni, R.~Thomale, R.~Comin, I.~Zeljkovic, J.~G.~Checkelsky, and S.~D.~Wilson, Kagome metals, Rev.~Mod.~Phys.~\textbf{98}, 015002 (2026)
\bibitem{moreno2022} J.~R.~Moreno, G.~Carleo, A.~Georges, and J.~Stokes, Fermionic wave functions from neural-network constrained hidden states, Proc.~Natl.~Acad.~Sci.~\textbf{119}, e2122059119
 (2022)
\bibitem{rende2026} R.~Rende, A.~Nikolaenko, L.~L.~Viteritti, S.~Sachdev, and Y.~-Hui Zhang, Transformer Neural-Network Quantum States for lattice models of spins and fermions: Application to the Ancilla Layer Model, arXiv:2603.02316 (2026)
\bibitem{viteritti2026} L.~L.~Viteritti, R.~Rende, C.~Roth, A.~Sengupta, G.~Carleo, and A.~Georges, Beyond Variational Bias: Resolving Intertwined Orders in the Hubbard Model, arXiv:2604.21978 (2026)

\end{thebibliography}
\end{document}